\documentclass[aps, two column,superscriptaddress, showpacs, amssymb]{revtex4-1}

\usepackage{graphicx} 
\usepackage{mathrsfs}
\usepackage{dcolumn}
\usepackage{bm}
\usepackage{amsmath}
\usepackage{booktabs}
\usepackage{multirow}
\pdfoutput=1

\begin{document}

\preprint{}

\title{The Optical Chirality Flux as a Useful Far-Field Probe of Chiral Near Fields}

\author{Lisa V. Poulikakos}
\affiliation{Optical Materials Engineering Laboratory, ETH Zurich, 8092 Zurich, Switzerland}
\author{Philipp Gutsche}
\affiliation{Zuse Institut Berlin, 14195 Berlin, Germany}
\author{Kevin M.  McPeak}\altaffiliation[Current address: ]{Louisiana State University, Baton Rouge, LA 70803, USA}
\affiliation{Optical Materials Engineering Laboratory, ETH Zurich, 8092 Zurich, Switzerland}
\author{Sven Burger}
\affiliation{Zuse Institut Berlin, 14195 Berlin, Germany}
\affiliation{JCMwave GmbH, 14050 Berlin, Germany} 
\author{Jens Niegemann}
\affiliation{Institute of Electromagnetic Fields, ETH Zurich, 8092 Zurich, Switzerland}
\author{Christian Hafner}
\affiliation{Institute of Electromagnetic Fields, ETH Zurich, 8092 Zurich, Switzerland}
\author{David J. Norris}\email{dnorris@ethz.ch}
\affiliation{Optical Materials Engineering Laboratory, ETH Zurich, 8092 Zurich, Switzerland}

\date{\today}

\begin{abstract}
To optimize the interaction between chiral matter and highly twisted light, quantities that can help characterize chiral electromagnetic fields near nanostructures are needed. Here, by analogy with Poynting's theorem, we formulate the time-averaged conservation law of optical chirality in lossy dispersive media and identify the optical chirality flux as an ideal far-field observable for characterizing chiral optical near fields. Bounded by the conservation law, we show that it provides precise information, unavailable from circular dichroism spectroscopy, on the magnitude and handedness of highly twisted fields near nanostructures.
\end{abstract}

\maketitle

\section{Introduction}

Chiral shapes are those not superimposable upon their mirror image. Chirality is common in nature, e.g., in biomolecules that exist in either left- or right-handed forms (known as enantiomers). It is also seen in knotted and twisted fields, such as fluid vortices \cite{moffatt1969degree, scheeler2014helicity}, magnetic flux tubes \cite{berger1999introduction}, and circularly polarized light \cite{kong1990electromagnetic}. Indeed, chiral electromagnetic fields have long been exploited to characterize chiral matter. More recently, their use has also been proposed for enantioselective biosensing \cite{tang2011enhanced, hendry2010ultrasensitive}, enantioselective separation \cite{canaguier2013mechanical, tkachenko2014optofluidic}, asymmetric catalysis \cite{takano2007asymmetric}, and nonlinear spectroscopic imaging \cite{rodrigues2014nonlinear, rodrigues2014metamaterials}. In many of these applications, metallic nanostructures are utilized because their local plasmonic resonances can create concentrated electromagnetic fields \cite{novotny2012principles}. When the nanostructures have a chiral shape, they can exhibit a high degree of optical activity \cite{gansel2009gold, hentschel2012three, fan2012chiral, mcpeak2014complex, hu2014self} and induce intense highly twisted chiral near fields \cite{schaferling2012tailoring, schaferling2014helical, mcpeak2015al}.

However, it is not always clear how to utilize these near fields most effectively. For example, in enantioselective biosensing, a molecule should interact with the near field of the chiral plasmonic nanostructure \cite{hendry2010ultrasensitive}. To maximize the effect, the near field should first be characterized through an appropriate far-field quantity. Typically, this has been done with circular dichroism (CD) spectroscopy, which records the difference in the optical extinction from a chiral object when it is illuminated with left- and right-handed circularly polarized light (for alternative methods see \cite{saba2013group, saba2015bloch}). For most biomolecules, the differential absorption dominates the CD signal and provides information about the chiral arrangement of the molecular components. For larger objects, differential scattering becomes increasingly important, yielding data about longer-range ($\agt$ 20 nm) structure \cite{bustamante1983circular}. While obtaining such knowledge is useful in many contexts, in applications such as enantioselective biosensing, we already know the nominal chiral shape of our plasmonic nanostructures. Rather, we wish to characterize the chirality of their intense near fields. This can be problematic with CD spectroscopy for two reasons: (i) excitation with circularly polarized light introduces local chirality in the system that is unrelated to the structure, and (ii) the differential extinction of left- and right-handed circularly polarized light can contain unwanted cancellation effects. Consequently, the amplitude and sign of the peaks in CD spectra do not directly relate to the chirality of the near field \cite{mcpeak2015al}. Thus, a need exists for a more useful physical quantity for characterizing chiral near fields.

Herein, we identify such a quantity, the optical chirality flux.  By analogy with Poynting's theorem, we formulate a time-averaged conservation law for chirality in lossy dispersive media. We show how a chiral nanostructure can selectively dissipate linearly polarized light and generate an optical chirality flux in the far field. Further, we verify both analytically and numerically that this quantity can then provide useful information, not obtainable from CD spectroscopy, about the amplitude and handedness of the near-field chirality of the nanostructure. Therefore, the optical chirality flux provides a useful far-field observable for studying chiral near fields.
\vspace{-0.5cm}
\section{Chirality Conservation}
Chiral electromagnetic fields exhibit a twist around an axis, which in general can be referred to as \textit{optical chirality}. More tightly wrapped field lines arise when a higher degree of local chirality exists \cite{tang2010optical}. This can be quantified via the optical chirality density \cite{philbin2013lipkin},
\begin{equation}
\chi = \frac{1}{2}[\textbf{D} \cdot (\nabla \times \textbf{E}) + \textbf{B} \cdot(\nabla \times \textbf{H})],  \label{chi_material}
\end{equation}
\noindent where $\textbf{E}$ and $\textbf{B}$ are the time-dependent electric and magnetic fields. We assume linear dispersive media such that $\textbf{D} = \epsilon(\omega) \textbf{E}$ and $\textbf{B} = \mu(\omega) \textbf{H}$ with $\epsilon = \epsilon' + i \epsilon''$ as the complex electric permittivity, $\mu = \mu' + i \mu''$ the complex magnetic permeability, and $\omega$ the angular frequency.  In its time-averaged form, Eq. (\ref{chi_material}) yields
\begin{equation}
\bar{\chi} = -\frac{\omega}{2} Im( \textbf{D}^* \cdot \textbf{B}),  \label{chi_time_avg}
\end{equation}
\noindent which is a time-even and parity-odd pseudoscalar \cite{lipkin1964existence}. 

Recently, $\chi$ has been interpreted as the degree of asymmetry in the excitation rate between a molecule and its enantiomer in a chiral local field \cite{tang2010optical}. Further, a significant increase in enantioselectivity was found when chiral molecules interact locally with electromagnetic fields with $\chi$ greater than in circularly polarized light \cite{tang2011enhanced}. Consequently, $\chi$ has since been used to study chiral plasmonic nanostructures \cite{hendry2010ultrasensitive, schaferling2012tailoring, meinzer2013probing, schaferling2014helical}. It has identified near fields of high optical chirality enhancement, defined as the time-averaged optical chirality density [Eq. (\ref{chi_time_avg})] normalized by the corresponding value for circularly polarized light \cite{schaferling2012tailoring}. It has also been used to calculate wavelength-dependent quantities by spatially averaging near a structure of interest \cite{garcia2013surface, yoo2014globally, schaferling2014helical} or by evaluating at a single spatial coordinate \cite{alizadeh2015plasmonically}. Although this can provide an indication of the spectral dependence of $\chi$ (e.g., for comparison with CD spectra \cite{garcia2013surface}), the results vary with choice of integration volume or spatial coordinate.

Instead of $\chi$, we focus on the conservation of optical chirality. If $\chi$ is combined with the optical chirality flux,
\begin{equation}
\mathbf{\Sigma} = \frac{1}{2}[\textbf{E} \times (\nabla \times \textbf{H}) - \textbf{H} \times (\nabla \times \textbf{E})], \label{sigma_def}
\end{equation}
\noindent a conservation law for $\chi$ analogous to Poynting's theorem can be obtained \cite{lipkin1964existence, bliokh2011characterizing, barnett2012duplex, tang2010optical}: 
\begin{equation}
\int_V \frac{\delta \chi}{\delta t} d^3x + \int_S \mathbf{\Sigma} \cdot \textbf{n} da = 0.
\label{conservation_law}
\end{equation}
Previously, such a law was defined for lossless dielectrics \cite{ragusa1992electromagnetic,  philbin2013lipkin}. We extend this treatment and consider lossy dispersive media by analogy with Poynting's theorem for time-harmonic fields. As shown below, this yields a physical interpretation of the dissipation of optical chirality and the optical chirality flux. In contrast to prior wavelength-dependent studies, the quantities obtained in this approach are uniquely defined and bounded by the conservation law.

For time-averaged, time-harmonic fields, Poynting's theorem for conservation of energy in source-free lossy dispersive media is \cite{jackson1999classical} 
\begin{equation}
-2 \omega \int_V Im(w_e - w_m) d^3x + \int_V Re(\nabla \cdot \mathcal{S}) d^3x = 0. 
\label{Poynting_medium_time_averaged}
\end{equation} 
\noindent
Here \textit{w$_e$} = $\frac{1}{4}(\mathcal{E} \cdot \mathcal{D}^*)$ and \textit{w$_m$} = $\frac{1}{4}(\mathcal{B} \cdot \mathcal{H}^*)$ are the harmonic electric- and magnetic-energy densities; $\mathcal{E}$, $\mathcal{D}$, $\mathcal{B}$, and $\mathcal{H}$ represent the complex time-harmonic fields; and $\mathcal{S}$ = $\frac{1}{2}(\mathcal{E} \times \mathcal{H}^*$) is the complex Poynting vector. The two terms in Eq. (\ref{Poynting_medium_time_averaged}) represent energy dissipation and energy flux, respectively. Because $Im(w_e -w_m) = -\frac{1}{4}(\epsilon'' |\mathcal{E}|^2+ \mu''  |\mathcal{H}|^2)$, the dissipation term is non-zero only for lossy media, i.e., those with imaginary components in the permittivity or permeability. 

In this work, we study source-free systems where $\textbf{J} = 0$. In general, the right-hand side of the time-domain conservation law of optical chirality [Eq. (\ref{conservation_law})] is $ -\frac{1}{2} [\textbf{J} \cdot (\nabla \times \textbf{E}) + \textbf{E} \cdot (\nabla \times \textbf{J})]$. Analogously to Poynting's theorem, by applying Maxwell's equations and vector identities to the time-averaged form of this expression (Appendix \ref{appA}), we obtain the time-averaged, time-harmonic conservation law for optical chirality in lossy dispersive media:
\begin{equation}
-2\omega \int_V Im(\chi_e - \chi_m) d^3x + \int_V Re(\nabla \cdot \mathscr{S} )d^3x = 0, \label{Chirality_conservation_medium_time_averaged}
\end{equation}
\noindent
where $\chi_e$ and $\chi_m$ are defined as the complex harmonic electric and magnetic optical chirality densities, and $\mathscr{S}$ is the corresponding optical chirality flux:
\begin{equation}
\chi_e = \frac{1}{8}[\mathcal{D}^* \cdot(\nabla \times \mathcal{E}) + \mathcal{E} \cdot (\nabla \times \mathcal{D}^*)],
\label{electric_chirality}
\end{equation}
\begin{equation}
\chi_m = \frac{1}{8}[\mathcal{H}^* \cdot(\nabla \times \mathcal{B}) + \mathcal{B} \cdot (\nabla \times \mathcal{H}^*)],
\label{magnetic_chirality}
\end{equation}
\begin{equation}
\mathscr{S} = \frac{1}{4}[\mathcal{E} \times(\nabla \times \mathcal{H}^*) - \mathcal{H}^* \times (\nabla \times \mathcal{E})]. 
\label{time_averaged_chirality_flux}
\end{equation}
\noindent

Equations (\ref{Chirality_conservation_medium_time_averaged}) to (\ref{time_averaged_chirality_flux}) form the central result of this work. This extension of the conservation law to lossy dispersive media identifies the mechanism of optical chirality dissipation upon interaction with matter [first term in Eq. (\ref{Chirality_conservation_medium_time_averaged})]. Furthermore, by studying its time-averaged, time-harmonic form, we establish the optical chirality flux [second term in Eq. (\ref{Chirality_conservation_medium_time_averaged})] as a physically measurable quantity. 

In general, the dissipation term in Eq. (\ref{Chirality_conservation_medium_time_averaged}) can be written as
\vspace{-0.1cm}
\begin{gather}
Im(\chi_e - \chi_m) = \nonumber \\ 
 \frac{1}{8}[-\nabla \epsilon' \cdot Im(\mathcal{E} \times \mathcal{E}^{*}) - \nabla \mu' \cdot Im(\mathcal{H} \times \mathcal{H}^*)] \\ 
+ \frac{1}{4}(\epsilon' \mu''+\epsilon''\mu')Im(\mathcal{E}^* \cdot \mathcal{H}) . \nonumber
\label{chirality_diss}
\end{gather}
\noindent
Equation (10) is then integrated over the entire structure. [In previous work, Eq. (\ref{chi_time_avg}) was evaluated in the space surrounding the structure.] The first term on the right hand side of Eq. (10) describes optical chirality dissipation via a gradient in $\epsilon'$ and $\mu'$. The examples presented here study piecewise homogeneous, isotropic media where $\nabla \epsilon(\omega)$ and $\nabla \mu(\omega)$ can only be non-zero between material domains. Thus, an interface is required for this dissipation mechanism \cite{gutscheinprep}. Note that the conservation law in Eq. (\ref{Chirality_conservation_medium_time_averaged}) is not restricted to piecewise homogeneous, isotropic media, but is generally applicable to anisotropic systems.
Additionally, the second term on the right hand side of Eq. (10) is non-zero only for lossy media ($\epsilon''$ $\neq$ 0 or $\mu''$ $\neq$ 0) in analogy to the dissipation term in Poynting's theorem [Eq. (\ref{Poynting_medium_time_averaged})].

\begin{figure}
\includegraphics{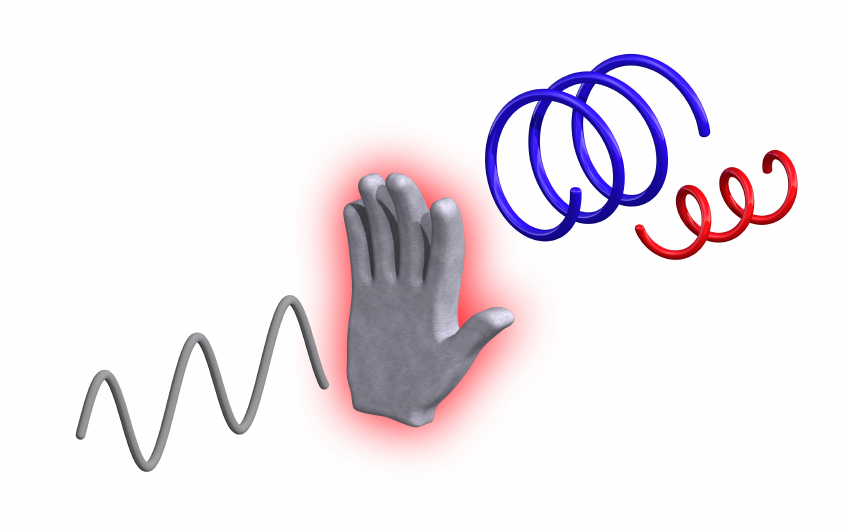}
\caption{\label{figure_1} Conservation of optical chirality for a chiral structure (hand) made of a lossy dispersive medium. The structure is excited (left) with achiral linearly polarized light. Due to the conservation law [Eq. (\ref{Chirality_conservation_medium_time_averaged})], dissipation of optical chirality in the structure causes an outgoing optical chirality flux (right).}
\end{figure}

Figure \ref{figure_1} summarizes these results schematically. A chiral structure (a hand) made of a homogeneous isotropic lossy dispersive medium is excited with vertically polarized light. As linear polarization is an equal superposition of left- and right-handed circular polarization, the incoming optical chirality flux is zero.  If the structure then selectively dissipates optical chirality of one handedness, a non-zero outgoing optical chirality flux must be generated according to Eq.  (\ref{Chirality_conservation_medium_time_averaged}). In this case, the only chiral light contained in the scattered field is due to the structure. Because the scattered optical chirality flux can be directly related to the polarization of the scattered light \cite{bliokh2011characterizing}, it is an easily measurable far-field quantity.

Indeed, for an electromagnetic plane wave $\mathscr{S}$ is proportional to the third Stokes parameter, an experimentally observable quantity describing the degree of circular polarization \cite{jackson1999classical}. The amplitude of an electric field can be written as $\textbf{A} = l \textbf{A}_{LCPL} + r \textbf{A}_{RCPL}$, where $l$ and $r$ are the weighting factors of left- and right-handed circular polarization. Thus, $\mathscr{S}$ is a linear combination of components of each handedness: $\mathscr{S} = |l|^2 \mathscr{S}_{LCPL} + |r|^2 \mathscr{S}_{RCPL}$ (analogous to the power flux $\mathcal{S}$). By applying Eq. (\ref{time_averaged_chirality_flux}) to left- and right-handed circularly polarized light (with $\mu$ = $\mu_{0}$), $\mathscr{S}$ can be directly related to $\mathcal{S}$:
\vspace{-0.1 cm}
\begin{gather}
\mathscr{S}_{LCPL} = \frac{\omega}{c}\mathcal{S}_{LCPL}, \\ \label{flux_LCPL}
\mathscr{S}_{RCPL} = -\frac{\omega}{c}\mathcal{S}_{RCPL}, \\
\mathscr{S} = \frac{\omega}{c}(|l|^2\mathcal{S}_{LCPL} - |r|^2\mathcal{S}_{RCPL}). \label{stokes_parameter}
\end{gather}
\noindent
Eq. (\ref{stokes_parameter}) is proportional to the third Stokes parameter.

Before proceeding, we briefly compare $\mathscr{S}$ to another quantity which is potentially useful for characterizing chiral fields, the optical helicity \cite{barnett2012duplex, fernandez2013electromagnetic, nieto2015optical}. Because the helicity depends on the electric vector potential, it is only uniquely defined for divergence-free $\textbf{D}$ fields. Further, although the optical helicity can be gauge invariant if integrated over all space \cite{barnett2012duplex}, it cannot be determined using only quantities obtained directly from Maxwell's equations. In contrast, the optical chirality flux can be calculated directly from the vector fields.

\section{Quasistatic Electric-Dipole Limit}\label{eldipole}
So far, we have identified the optical chirality flux as a uniquely defined, far-field observable. We now consider whether it can provide information about the magnitude and handedness of the time-averaged optical chirality density [Eq. (\ref{chi_time_avg})] in the near field surrounding a chiral structure. For this, we first treat a small sphere in the quasistatic electric-dipole limit analytically. 

In this system, both the local electric-field intensity and the scattered power exhibit a resonance due to the denominator in the electric-dipole polarizability, $\alpha_e = 4 \pi \epsilon_0 a^3 \{\epsilon_1(\omega)-\epsilon_2\}/\{\epsilon_1(\omega)+2\epsilon_2\}$. Here $a$ is the sphere radius, and $\epsilon_0$, $\epsilon_1$, and $\epsilon_2$ are the permittivity in vacuum, the structure, and the surrounding medium, respectively \cite{novotny2012principles}. 
Because the sphere is achiral, we introduce chirality to this system by exciting it with circularly polarized light. 

Specifically, we study the electric field $\textbf{E}_{tot} = \textbf{E}_{inc} + \textbf{E}_{scat}$ and the magnetic field $\textbf{H}_{tot} = \textbf{H}_{inc} + \textbf{H}_{scat} \approx \textbf{H}_{inc}$, where the incident field corresponds to that of left-handed circularly polarized light propagating in +z. Furthermore, in the quasistatic limit, where the near zone of an electric dipole 
extends to infinity, the scattered fields are defined as the static dipole fields with harmonic time dependence \cite{jackson1999classical}. 

The optical chirality density $\bar{\chi}$ [Eq. (\ref{chi_time_avg})] and optical chirality flux $\mathscr{S}$ [Eq. (\ref{time_averaged_chirality_flux})] are then:
\begin{equation}
\bar{\chi}_d = \frac{E_0^2 k}{16 \pi r^3} \Big\{16 \pi r^3 \epsilon_0 - Re(\alpha_{e})[1+3 cos2\theta] \Big\}, \label{chi_dipole}
\end{equation}
\begin{equation}
\mathscr{S}_d= \frac{\omega}{2k}  \bar{\chi}_d 
\left(
\begin{array}{c}
cos\theta\\
-sin\theta\\
0 \end{array}
\right),
\label{sigma_dipole}
\end{equation}

\begin{figure}
\includegraphics{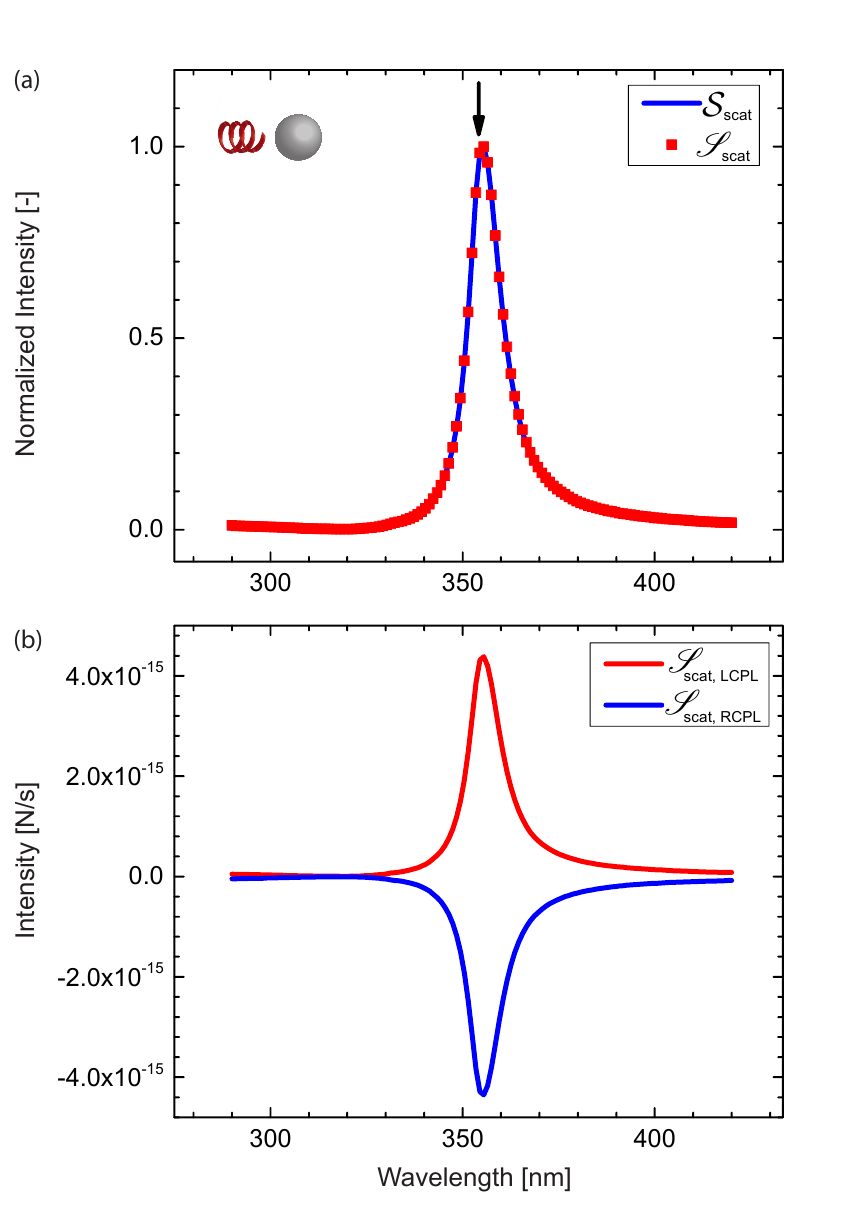}
\caption{\label{figure_2} (a) The scattered power, $\mathcal{S}_{scat}$, and scattered optical chirality flux, $\mathscr{S}_{scat}$, for a 10-nm-diameter Ag sphere excited with left-handed circularly polarized light (inset). Both quantities are integrated over the sphere volume and normalized. Their spectra match exactly according to the resonance in the electric-dipole polarizability $\alpha_e$ (vertical arrow). The peak in $\mathscr{S}_{scat}$ corresponds to the near-field resonance of the optical chirality density calculated analytically in the electric-dipole limit.
(b) Optical chirality flux, $\mathscr{S}_{scat}$, for a 10-nm-diameter Ag nanosphere, upon excitation with left- ($\mathscr{S}_{scat, LCPL}$) and right-handed ($\mathscr{S}_{scat, RCPL}$) circularly polarized light. The sign of $\mathscr{S}_{scat}$ changes with the handedness of the excitation source.}
\end{figure}
\noindent
where $E_0$ is the amplitude of the incoming electric field, $k$ the wavenumber,  $r$ the radial coordinate, and $\theta$ the polar angle. Equations (\ref{chi_dipole}) and (\ref{sigma_dipole}) describe the spatial dependence of the optical chirality density and optical chirality flux in this system. 

Equation (\ref{chi_dipole}) contains two terms, corresponding to the optical chirality density of the incoming and scattered fields, respectively. The same two terms are in the optical chirality flux [Eq. (\ref{sigma_dipole})]. For the scattered field, this means that both the optical chirality density and flux exhibit the same resonance condition, defined by the real part of $\alpha_{e}$. This corresponds to the Fr{\"o}hlich condition where Im($\epsilon_1$) is small or slowly varying at resonance \cite{maier2007plasmonics} (Appendix \ref{el_dipole}).

Figure \ref{figure_2}(a) verifies this analytical treatment via numerical simulations (Appendix \ref{FEM}) of a 10-nm-diameter Ag nanosphere. The scattered power and scattered optical chirality flux are integrated over a volume slightly larger than the nanosphere to yield $\mathcal{S}_{scat}$ and $\mathscr{S}_{scat}$, respectively. We note that for optical chirality, magnetic dipolar contributions must also be considered, as discussed in Appendix \ref{elmagdipole}. As expected, a peak appears in the spectrum of each quantity at the resonance in $Re(\alpha_{e})$. At this frequency, a near-field resonance also occurs in the optical chirality density [Eq. (\ref{chi_dipole})]. Thus, a direct connection exists between the optical chirality density in the near field of a structure and the optical chirality flux in the far field.

In Fig. \ref{figure_2}(b) we show that excitation with left- or right-handed circularly polarized light results in mirror symmetric $\mathscr{S}_{scat}$ spectra for an achiral nanosphere. This is expected from the chirality conservation law because the sphere is excited with an incoming optical chirality flux of positive or negative sign, respectively. 

\section{Chiral Plasmonic Nanorod Dimer}\label{sec_NR}
\begin{figure}
\includegraphics{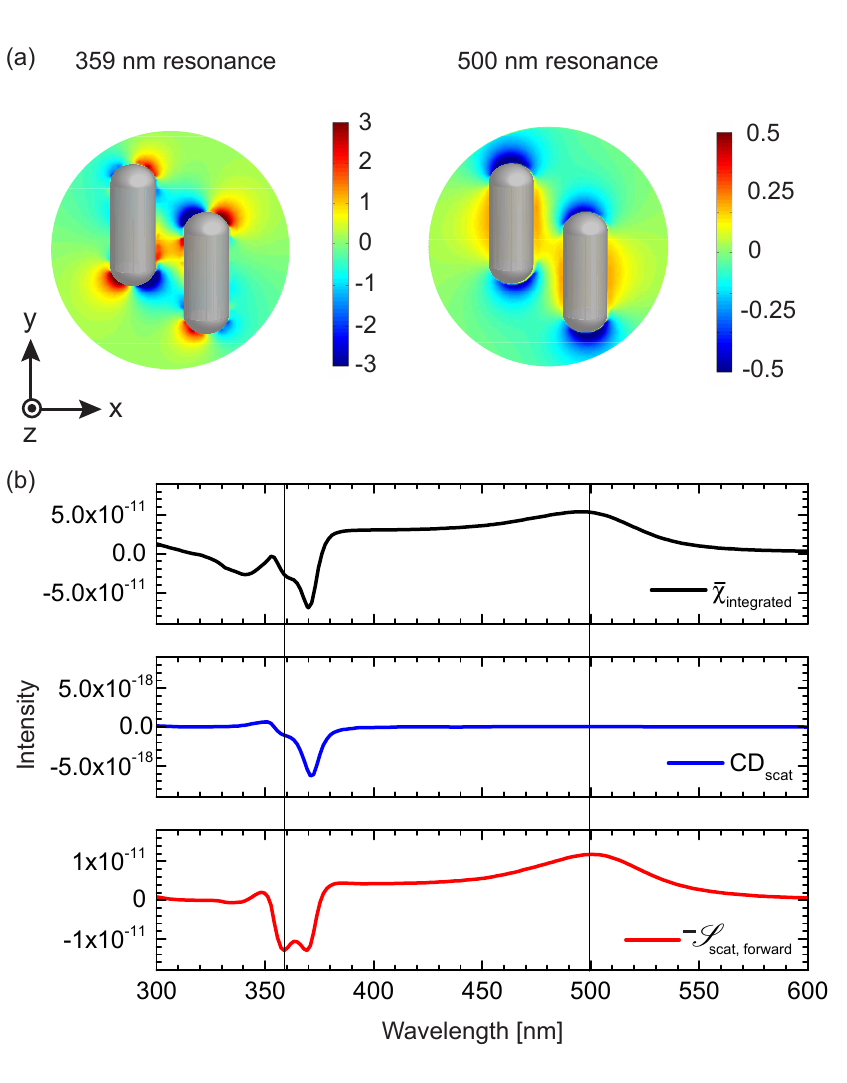}
\caption{\label{figure_3} (a) Near-field maps (z\,=\,0 plane) of the optical chirality enhancement of a Ag nanorod dimer with hemispherically-capped cyllindrical nanorods (40 nm diameter, 80 nm length, 20 nm spacing). The right nanorod is shifted 40 nm in the -y direction to create a two-dimensionally chiral structure. The dimer is excited with x-polarized light propagating along +z at a transverse (359 nm) and longitudinal (500 nm) resonance. (Further transverse resonances at 347 nm and 369 nm are shown in Appendix \ref{NR_dimer}.)  The transverse (longitudinal) resonance shows primarily positive (negative) optical chirality density. (b) \textit{Top:} integrated optical chirality density [Eq. (\ref{chi_time_avg})] in newtons. \textit{Center:} scattered component of the dimer's CD spectrum in watts. \textit{Bottom:} forward-scattered optical chirality flux in N/s. For comparison, $-\mathscr{S}_{scat,forward}$ is plotted. 
The dominant sign of the field maps in (a) is also seen in the optical chirality flux. 
In the CD spectrum, the longitudinal resonance is absent.}
\end{figure}

Due to the connection between optical chirality density in the near field and optical chirality flux in the far field of a structure, one might conclude that these quantities are equivalent for characterizing the chiral near field. However, $\mathscr{S}$ is clearly superior because it is bounded by the conservation law. We confirm this with a specific example. For simplicity, we treat a Ag nanorod dimer that exhibits only two-dimensional chirality (see Fig. \ref{figure_3}). With finite-element simulations (Appendix \ref{FEM}), we excite the structure with x-polarized light propagating along +z. Figure \ref{figure_3}(a) maps the optical chirality enhancement of the dimer at a transverse (359 nm) and longitudinal (500 nm) resonance for a slice in the x-y plane at z\,=\,0 (the center of the rods). (Further transverse resonances at 347 nm and 369 nm are shown in Appendix \ref{NR_dimer}.) The three panels of Fig. \ref{figure_3}(b) plot: the optical chirality density [Eq. (\ref{chi_time_avg})] integrated over the space near the dimer, the scattered component of the dimer's CD spectrum (again excited along +z), and the optical chirality flux scattered in the forward direction. We chose this last parameter because the polarization of forward scattered light can be detected in an experimental transmission measurement. Due to the conservation law [Eq. (\ref{Chirality_conservation_medium_time_averaged})], the optical chirality flux can be determined with a surface integral of Eq.  (\ref{time_averaged_chirality_flux}) over the boundaries of the computational domain, following Gauss' law. Thus, unlike the volume integral of the optical chirality density [top panel in Fig. \ref{figure_3}(b)], the optical chirality flux remains constant for any computational domain containing the entire structure (Appendix \ref{NR_dimer}).

According to the plots in Fig. \ref{figure_3}(a), the transverse resonance exhibits primarily positive (left-handed) optical chirality density, while the longitudinal resonance is primarily negative (right-handed). This change of sign is also seen in the integrated optical chirality density and flux [top and bottom panels of Fig. \ref{figure_3}(b)]. However, the integrated optical chirality density (top panel) does not match the polarity of the sign in Fig. \ref{figure_3}(a), i.e., positive at 359 and negative at 500 nm. Indeed, as expected, we found that the shape and sign of the $\bar{\chi}$ spectrum depended on the integration volume chosen.

\begin{figure}
\includegraphics[width=\columnwidth]{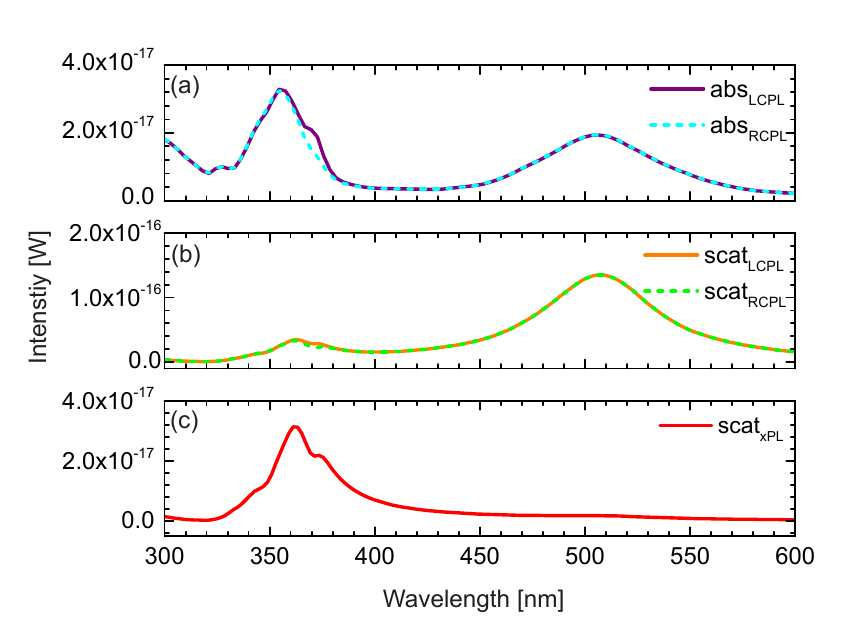}
\caption{(a) Absorption spectra upon excitation with left- ($abs_{LCPL}$) and right-handed circularly polarized light  ($abs_{RCPL}$). (b) Scattering spectra upon excitation with left- ($scat_{LCPL}$) and right-handed circularly polarized light $(scat_{RCPL}$). (c) Scattered power spectrum ($scat_{xPL}$) upon excitation with x-polarized light.}
\label{figure_4}
\end{figure}

Meanwhile, the CD spectrum (center panel) does not even exhibit a resonance at 500 nm. The signals at this longitudinal resonance for left- and right-handed circularly polarized light are equal and cancel. This effect is confirmed in Figs. \ref{figure_4}(a) and (b), where the absorbed and scattered spectra are shown for left- and right-handed circularly polarized light separately. The CD spectrum results from the difference between these two components, which are identical at long wavelengths. Similar cancellation effects have been observed for chiral plasmonic nanorod dimers in prior work \cite{auguie2011fingers} and can be explained within the coupled dipole oscillator model \cite{fan2010plasmonic, norden1997circular}. [For comparison, the scattered power spectrum upon x-polarized light excitation, $scat_{xPL}$, is shown in Fig. \ref{figure_4}(c).]

Figures \ref{figure_4}(a) and (b) also show that the largest differential absorption and scattering occurs at 369 nm, leading to a large CD amplitude. However, due to cancellation effects, this differential component is not necessarily related to the optical chirality density in the near field of the chiral nanorod dimer. Further details on the CD spectra are discussed in Appendix \ref{NR_dimer}.

Finally, we note that a non-zero optical chirality flux represents a dominance of either left- or right-handed optical chirality that is scattered into the far field. High optical chirality flux can occur even for a resonance where the local near-field enhancement is small. The nanorod dimer displays such a resonance at 500 nm, as indicated by the high magnitude in $\mathscr{S}_{scat}$ despite low values in the optical chirality enhancement [Fig. \ref{figure_3}(a)] and scattered power [Fig. \ref{figure_4}(c)]. Finding such resonances can be extremely useful (e.g., in enantioselective separation \cite{canaguier2013mechanical}) and is not possible via the local optical chirality density, which scales with the field enhancement \cite{canaguier2014chiral}. 
\vspace{-0.2 cm}
\section{Conclusion}
In summary, we have formulated the conservation law of optical chirality in its time-averaged form for lossy dispersive media. We identify the optical chirality flux as a physically useful observable that provides information on the magnitude and dominant handedness of the optical chirality in the near field of a structure. The same information cannot be obtained with conventional CD spectroscopy. Thus, the optical chirality flux can be exploited as a far-field probe of local resonances at which one handedness dominates the chiral electromagnetic fields. This can allow more effective use of chiral plasmonic nanostructures in enantioselective applications.

\begin{acknowledgments}
We thank C. Genet, A. Canaguier-Durand, M. Sch{\"a}ferling, E. De Leo, F. Prins, B. le Feber, and F. Schmidt for helpful discussions.  This work was supported by the Swiss National Science Foundation (Award no. 200021-146747) and the Einstein Foundation Berlin (ECMath, project SE6). Computations were performed at the ETH High-Performance Computing Cluster Brutus.
\end{acknowledgments}
\appendix

\section{Derivation of the Time-Averaged Conservation Law for Optical Chirality in Lossy Dispersive Media}\label{appA}

\renewcommand\theequation{A\arabic{equation}}
\setcounter{subsection}{0}
\setcounter{equation}{0}

For the time-harmonic form of a field quantity $\textbf{X}$, we use the notation $\textbf{X}$ = $Re(\mathcal{X} e^{-i \omega t})$ where $\mathcal{X}$ represents the time-harmonic vector field. Thus, Maxwell's equations in complex, time-harmonic fields can be written as:
\begin{gather}
\nabla \times \mathcal{E} = i \omega \mathcal{B}  \label{maxwell_1}, \\
\nabla \times \mathcal{H} = - i \omega \mathcal{D} + \mathcal{J}  \label{maxwell_2},
\end{gather}
\noindent
where we assume time-independent, linear, dispersive media such that $\mathcal{D} = \epsilon(\omega) \mathcal{E}$ and $\mathcal{B} = \mu(\omega) \mathcal{H}$. Furthermore, $\nabla \cdot \mathcal{B} = 0$ and $\nabla \cdot \mathcal{D} = \rho$, where $\rho$ represents the electric-charge density.   
The derivation of the conservation law is shown in Table \ref{t1} and applies Maxwell's equations and vector identities by analogy with the time-averaged form of Poynting's theorem \cite{jackson1999classical} for lossy dispersive media. In Table \ref{t1}, rows (R2) and (R6) apply Maxwell's equations while rows (R3), (R4) and (R5) use appropriate vector identities. 
The coefficients of the quantities defined in Eqs. (\ref{electric_chirality}) - (\ref{time_averaged_chirality_flux}) can be understood from rows (R7) and (R8).

\begin{table}[h!]
	\caption{
Derivation of the time-averaged conservation law for optical chirality in lossy dispersive media.}
	\label{t1}
     	$\begin{array}{l l}
	\toprule
	\hline
	\hline
	\textrm{Row} & \vline  \textrm{Chirality Conservation} \\
	\hline
	\midrule
	\addlinespace
\begin{aligned}
			(R1)
		\end{aligned} &
		\vline
		\begin{aligned}
			\mathcal{J}^* \cdot (\nabla \times \mathcal{E}) + \mathcal{E} \cdot (\nabla \times \mathcal{J}^*)  \\ 
		\end{aligned} \\ 
	\addlinespace
	\hline
	\midrule
	\addlinespace
		\begin{aligned}
			(R2)
		\end{aligned} &
		\vline
		\begin{aligned}
			= (\nabla \times \mathcal{H}^*) \cdot (\nabla \times \mathcal{E}) - i \omega \mathcal{D}^* \cdot (\nabla \times \mathcal{E}) \\
			+ \mathcal{E} \cdot [\nabla \times (\nabla \times \mathcal{H}^*)] - i \omega \mathcal{E} \cdot (\nabla \times \mathcal{D}^*)
		\end{aligned} \\ 
	\addlinespace
	\hline
	\midrule
	\addlinespace
		\begin{aligned}
			(R3)
		\end{aligned} &
		\vline
		\begin{aligned}
			= \mathcal{H}^* \cdot [\nabla \times (\nabla \times \mathcal{E})] - i \omega \mathcal{D}^* \cdot (\nabla \times \mathcal{E}) \\
			+\mathcal{E} \cdot [\nabla \times(\nabla \times \mathcal{H}^*)] \\
			- i \omega \mathcal{E} \cdot (\nabla \times \mathcal{D}^*) - \nabla \cdot [(\nabla \times \mathcal{E}) \times \mathcal{H}^*]
		\end{aligned} \\ 
	\addlinespace
	\hline
	\midrule
	\addlinespace
		\begin{aligned}
			(R4)
		\end{aligned} &
		\vline
		\begin{aligned}
			= \mathcal{H}^* \cdot [\nabla \times (\nabla \times \mathcal{E})] - i \omega \mathcal{D}^* \cdot (\nabla \times \mathcal{E}) \\
			+ (\nabla \times \mathcal{H}^*) \cdot (\nabla \times \mathcal{E}) - i \omega \mathcal{E} \cdot (\nabla \times \mathcal{D}^*) \\
			- \nabla \cdot [(\nabla \times \mathcal{E}) \times \mathcal{H}^*] - \nabla \cdot [\mathcal{E} \times (\nabla \times \mathcal{H}^*)]
		\end{aligned}\\ 
	\addlinespace
	\hline
	\midrule
	\addlinespace
		\begin{aligned}
			(R5)
		\end{aligned} &
		\vline
		\vline
		\begin{aligned}
			= \mathcal{H}^* \cdot [\nabla \times (\nabla \times \mathcal{E})] - i \omega \mathcal{D}^* \cdot (\nabla \times \mathcal{E}) \\
			+ (\nabla \times \mathcal{H}^*) \cdot (\nabla \times \mathcal{E}) - i \omega \mathcal{E} \cdot (\nabla \times \mathcal{D}^*) \\
			- \nabla \cdot [\mathcal{E} \times (\nabla \times \mathcal{H}^*) - \mathcal{H}^* \times (\nabla \times \mathcal{E})]
		\end{aligned}\\ 
	\addlinespace
	\hline
	\midrule
	\addlinespace
		\begin{aligned}
			(R6)
		\end{aligned} &
		\vline
		\vline
		\begin{aligned}
			= i \omega \mathcal{H}^* \cdot [\nabla \times \mathcal{B}] - i \omega \mathcal{D} ^* \cdot ( \nabla \times \mathcal{E}) \\
			+ i \omega ( \nabla \times \mathcal{H}^*) \cdot \mathcal{B} - i \omega \mathcal{E} \cdot (\nabla \times \mathcal{D}^*) \\
			- \nabla \cdot [ \mathcal{E} \times (\nabla \times \mathcal{H}^*) - \mathcal{H}^* \times (\nabla \times \mathcal{E})]
		\end{aligned}\\
	\addlinespace
	\hline
	\midrule
	\addlinespace
		\begin{aligned}
			(R7)
		\end{aligned} &
		\vline
		\vline
		\begin{aligned}
			= 8 i \omega \chi_m - 8 i \omega \chi_e - \nabla \cdot 4 \mathscr{S}
		\end{aligned}\\
	\addlinespace
	\hline
	\hline
	\midrule
	\addlinespace
	\addlinespace
		\begin{aligned}
			(R8)
		\end{aligned} &
		\vline
		\vline
		\begin{aligned}
			\addlinespace
			 -\frac{1}{4}[\mathcal{J}^* \cdot (\nabla \times \mathcal{E}) + \mathcal{E} \cdot (\nabla \times \mathcal{J}^*)]  \\
			= 2 i \omega (\chi_e - \chi_m) + \nabla \cdot \mathscr{S}
		\end{aligned}\\
	\addlinespace
	\hline
	\hline
	\bottomrule
     	\end{array}$
\end{table}                       

We implement the chirality conservation law in our numerical calculations for piecewise homogeneous, isotropic media, where the the electric and magnetic optical chirality densities and the optical chirality flux reduce to
\begin{gather}
\chi_e = \frac{1}{8} i \omega [ \mathcal{D}^* \cdot \mathcal{B} - (\epsilon \mathcal{B})^* \cdot \mathcal{E}],\label{chi_e_hom} \\
\chi_m = \frac{1}{8} i \omega [\mathcal{D}^* \cdot \mathcal{B} - \mathcal{H}^* \cdot (\mu \mathcal{D})], \label{chi_m_hom} \\
\mathscr{S} = \frac{1}{4} i \omega (\mathcal{E} \times \mathcal{D}^* - \mathcal{H}^* \times \mathcal{B}),  
\end{gather}

\noindent
within a given material domain. 
We note that within a homogeneous, isotropic medium the time-averaged optical chirality density $\bar{\chi}$ [Eq. (2)] can be expressed as the real part of the sum of the electric and magnetic chirality densities [Eqs. (\ref{chi_e_hom}) and (\ref{chi_m_hom})]: $\bar{\chi} = Re(\chi_e + \chi_m)$.  

\section{The Fr{\"o}hlich Condition in the Quasistatic Electric-Dipole Limit}
\label{el_dipole}

\renewcommand\theequation{B\arabic{equation}}
\setcounter{subsection}{0}
\setcounter{equation}{0}

In this section, we compare the resonance condition for electric-field enhancement and scattered power \cite{novotny2012principles} with that of optical chirality density and scattered optical chirality flux. While in the former case, the resonance condition is defined by $\alpha_e \propto \frac{\epsilon_1(\omega) - \epsilon_2}{\epsilon_1(\omega)+2\epsilon_2}$, the resonance condition in the latter case [see Eq. \ref{chi_dipole}] is defined by the real part of the electric dipole polarizability $Re(\alpha_{e}) \propto \frac{(\epsilon_1'-\epsilon_2)(\epsilon_1'+2 \epsilon_2) + \epsilon_1^{''2}}{(\epsilon_1'+2\epsilon_2)^2+\epsilon_1^{''2}}$, where $\epsilon_1 = \epsilon_1'+ i\epsilon_1''$ for a lossy, dispersive medium. The Fr{\"o}hlich condition states that for small or slowly varying $\epsilon_1''$ in the resonant wavelength range, the resonance condition of $\alpha_e$ occurs at $\epsilon_1' = -2\epsilon_2$ \cite{maier2007plasmonics}. In this regime, both $\alpha_e$ and $Re(\alpha_{e})$ follow approximately the same resonance condition. This is valid for the permittivity of Ag \cite{johnson1972optical}, which is studied in this work and explains the matching resonance conditions for the quantities shown in Fig. \ref{figure_2}(a).

\renewcommand\theequation{C\arabic{equation}}
\setcounter{subsection}{0}
\setcounter{equation}{0}

\section{Finite-Element Simulations}\label{FEM}
The numerical computations performed in this work were conducted using the finite-element solver JCMsuite (JCMwave, Germany) and resulted in time-harmonic, electromagnetic near-fields governed by Maxwell's equations. Data from Johnson and Christy \cite{johnson1972optical} were used for the optical material constants of the studied Ag nanostructures embedded in vacuum. 

Spectral information was obtained with wavelength scans at a resolution of 1.5 nm. The structures were excited with a circularly or linearly polarized plane wave propagating in +z. Quantitative information on the scattered power, the CD spectrum, the scattered optical chirality flux, and the integrated optical chirality density was obtained from the electromagnetic near fields by surface or volume integration of the electric-field energy density and flux or the optical chirality density and flux over the boundary of the computational domain.

For the nanosphere, a spherical computational domain (20 nm diameter) was chosen, while for the chiral nanorod dimer we applied a cylindrical computational domain. For spectral information on the reflected and transmitted components of the scattered power and optical chirality flux, we divided this cylindrical computational domain into two parts through the center (each cylinder with 50 nm height, 100 nm diameter). For the spectral information on the integrated optical chirality flux, a single, compact cylindrical domain was applied (60 nm height, 100 nm diameter). The computational domains were bounded by a perfectly matched layer (PML).

For the nanorod dimer, the spatially-dependent optical chirality density at the resonant wavelengths was calculated in MATLAB from the electric- and magnetic-field data obtained in JCMsuite. A 4 x 4 median filter was used to remove numerical noise which can occur at sharp interfaces between mesh elements.

Both the finite-element polynomial degree and the side-length constraint of the mesh were chosen such that the relative numerical discretization error resulted in values $< 5 \%$ for the quantities calculated in the significant wavelength ranges. 

\renewcommand\theequation{D\arabic{equation}}
\setcounter{subsection}{0}
\setcounter{equation}{0}
\section{The Quasistatic Electric-Magnetic Dipole Limit} \label{elmagdipole}

Section \ref{eldipole} discusses the quasistatic electric-dipole limit as a simple analytical example to illustrate the connection between optical chirality density and optical chirality flux. Although small metallic nanospheres ($d \ll \lambda$) contain both electric and magnetic dipolar components, expressed by their respective polarizabilities, $\alpha_e$ and $\alpha_m$, the electric-dipole limit typically serves as an accurate approximation when calculating quantities such as the electric-field enhancement or scattered power because $\alpha_m \ll \alpha_e$ \cite{capolino2009theory}. 
However, we find that when computing the optical chirality density and flux for a metallic nanosphere, both electric and magnetic dipolar components must be considered. We show this by integrating the expressions for $\bar{\chi}_d$ [Eq. (\ref{chi_dipole})] and $\mathscr{S}_d$ [Eq. (\ref{sigma_dipole})] in the electric-dipole limit over the polar and azimuthal angles of a spherical surface:
\vspace{-0.2 cm}
\begin{gather}
\int_0^{2 \pi}\int_0^{\pi} \bar{\chi}_{d} (r^2 sin\theta) d\theta d\phi = 4 \pi r^2 \epsilon_0 E_0^2 k \label{chi_int}, \\
\int_0^{2 \pi}\int_0^{\pi} \mathscr{S}_{d} \cdot \vec{n} (r^2 sin\theta) d\theta d\phi = 0 \label{sigma_int}.
\end{gather}

\noindent
Here, Eq. (\ref{chi_int}) shows that only the optical chirality density of the incoming light remains in the system while Eq. (\ref{sigma_int}) shows that the integrated optical chirality flux is zero in the quasistatic electric-dipole limit.
Due to the chirality conservation law [Eq. (\ref{Chirality_conservation_medium_time_averaged})], in a system where no optical chirality dissipation occurs, the incoming and outgoing optical chirality flux are equal and the surface integral in Eq. (\ref{sigma_int}) is zero. This is the case of the electric-dipole limit.  However, in our numerical calculations of the optical chirality flux, $\mathscr{S}$, for a spherical Ag nanoparticle (Fig. \ref{figure_2}), we find that the magnetic dipolar component does indeed contribute to optical chirality dissipation and the generation of a net optical chirality flux. Thus, as opposed to calculations of quantities such as electric-field enhancement and scattered power, we must go beyond the electric-dipole limit to accurately predict the optical chirality flux scattered from a metallic nanoparticle.

Consequently, we studied the quasistatic electric-magnetic dipole limit to accurately predict the behavior observed for a spherical Ag nanoparticle in Fig. \ref{figure_2}. We neglected electric fields produced by a magnetic dipole and magnetic fields produced by an electric dipole and assumed that bi-anisotropy is not generated when both electric and magnetic dipoles are located at the same spatial position \cite{garcia2013surface}. The incident fields were defined as stated in Sec. \ref{eldipole}. However, as opposed to the electric-dipole limit, here the scattered components of the magnetic field are non-zero and our total magnetic field becomes $ \textbf{H}_{tot} = \textbf{H}_{inc} + \textbf{H}_{scat}$ where $\textbf{H}_{scat} = \frac{1}{4\pi}[3 \textbf{n}(\textbf{n} \cdot \textbf{p}_m)-\textbf{p}_m] \frac{1}{r^3}$ with the magnetic-dipole moment, $\textbf{p}_m = \alpha_m \textbf{H}_{inc}$, and magnetic-dipole polarizability, $\alpha_{m} = Re(\alpha_{m}) + i Im(\alpha_{m})$. 
Analytical calculations of the optical chirality density [Eq. (\ref{chi_time_avg})] in the space surrounding the electric-magnetic dipole resulted in:
\vspace{-0.3 cm}
\begin{gather}
\bar{\chi}_{el-mag} =  \\
\frac{E_0^2 k}{16 \pi r^3}\Big\{16 \pi r^3 \epsilon_0
-[Re(\alpha_e) + \epsilon_0 Re(\alpha_m)][1+3cos(2\theta)] \nonumber  \\
+ \frac{1}{4 \pi r^3}[Re(\alpha_e)Re(\alpha_m)
+ Im(\alpha_e)Im(\alpha_m)] [7-3cos(2\theta)] \Big\}. \nonumber
\label{chi_el_mag_eqn}
\end{gather}

The first term in Eq. (\ref{chi_el_mag_eqn}3) corresponds to the contribution of the incoming light, the second set of terms corresponds to the contribution of an electric and magnetic dipole on its own [compare Eq. (\ref{chi_dipole})], and the third set of terms represents the additional contribution when both electric and magnetic dipolar components are present. When $\bar{\chi}_{el-mag}$ is integrated over a spherical surface, only the first and third set of terms in Eq. (\ref{chi_el_mag_eqn}3) yield a non-zero value.
The additional third set of terms shows that for the optical chirality of a metallic nanosphere, both electric and magnetic dipolar contributions must be considered. Further, for the example of a spherical Ag nanoparticle (Fig. 2), $\alpha_m \ll \alpha_e$ and Im($\alpha_e$) is small or slowly varying at resonance. Thus, Re($\alpha_e$) remains the dominant term defining the resonance condition of $\bar{\chi}_{el-mag}$.

\renewcommand\theequation{E\arabic{equation}}
\setcounter{subsection}{0}
\setcounter{equation}{0}

\section{Chiral Plasmonic Nanorod Dimer} \label{NR_dimer}
In this section, we provide further details on the computations performed for the Ag chiral nanorod dimer. 
\vspace{-0.4 cm}
\subsection{Effect of Computational Domain}
\begin{figure}
\includegraphics{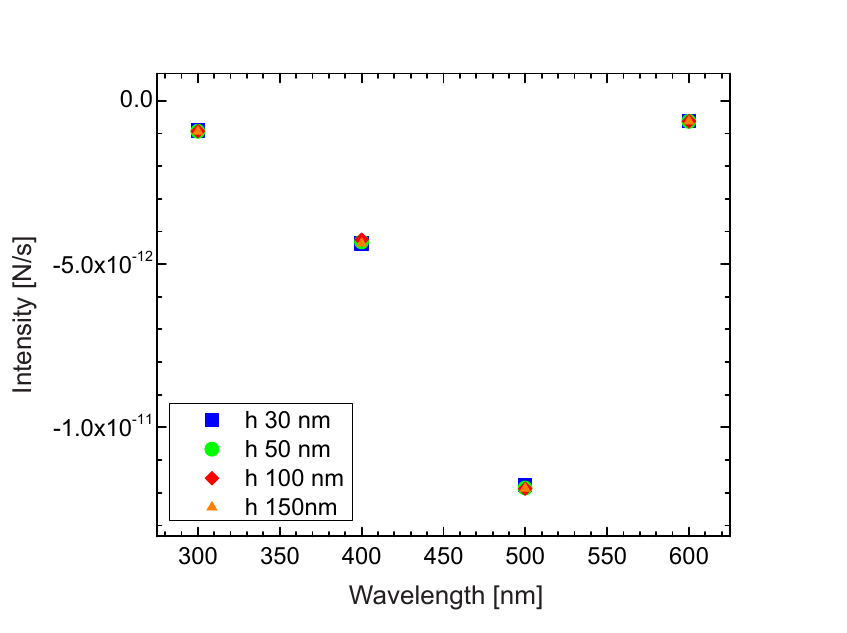}
\caption{Optical chirality flux in the forward direction, $\mathscr{S}_{scat}$, for cylindrical computational domains of 100 nm diameter but various heights in the +z-direction, h = 30, 50, 100, and 150 nm.} 
\label{size_comp_domain}
\end{figure}
The effect of the size of the computational domain on the scattered optical chirality flux, $\mathscr{S}_{scat}$, is shown in Fig. \ref{size_comp_domain} at four wavelengths in the relevant range. Because this quantity is bounded by the conservation law of optical chirality, its value is independent of the chosen size of computational domain. This is a major advantage in comparison to quantities such as the optical chirality density [Eq. (\ref{chi_time_avg})] that yield integrated values that depend on the choice of computational domain.
\vspace{-0.3 cm}

\subsection{Transverse Resonances}
\begin{figure}
\includegraphics{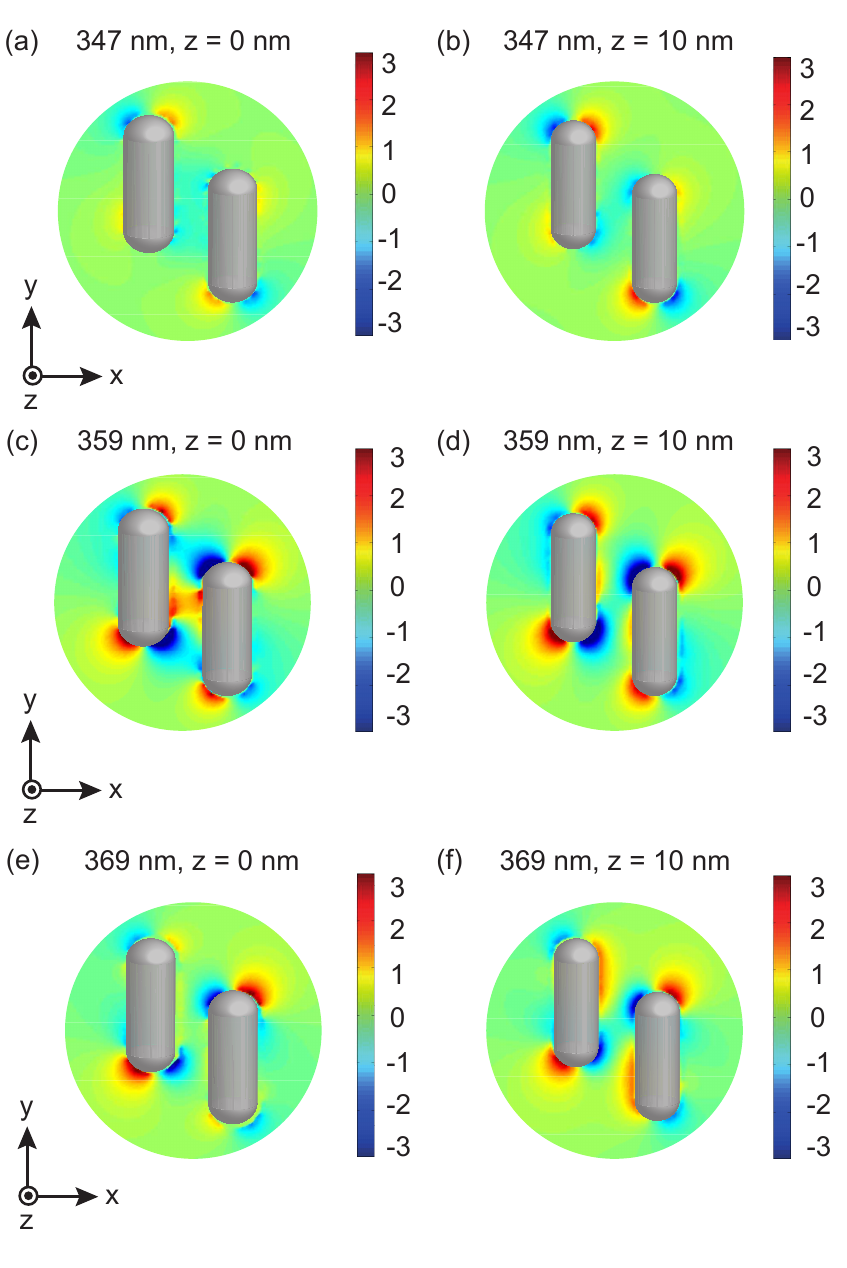}
\caption{Near-field maps of the optical chirality enhancement of the Ag nanorod dimer (see Fig. \ref{figure_3}) at the three transverse resonances at 347 nm [(a) and (b)], 359 nm [(c) and (d)] and 369 nm [(e) and (f)]. The maps display the x-y plane at z = 0 nm [(a), (c) and (d)] and at z = 10 nm [(b), (d) and (f)].} 
\label{transverse_res}
\end{figure}

In Fig. \ref{transverse_res} we show near-field maps of the optical chirality enhancement for the three transverse resonances exhibited by the chiral nanorod dimer (see Fig. \ref{figure_3}). Figures \ref{transverse_res}(a) and (b) display the first transverse resonance (347 nm) while Figs. \ref{transverse_res}(c) and (d) show the second (359 nm) and Figs. \ref{transverse_res}(e) and (f) the third transverse resonance (369 nm) respectively. The x-y plane at z = 0 nm (the center of the rods) is shown in Figs. \ref{transverse_res}(a), (c) and (e) while maps at z = 10 nm are shown in Figs. \ref{transverse_res}(b), (d) and (f). 

The optical chirality flux scattered in the forward direction, $\mathscr{S}_{scat, forward}$ [bottom panel of Fig. \ref{figure_3}(b)], exhibits positive extrema of similar amplitude at the second (359 nm) and third (369 nm) transverse resonances. Thus, $\mathscr{S}_{scat, forward}$ predicts a dominance of positive, left-handed optical chirality density in the near field in both cases. Indeed, this is confirmed by the near-field maps shown in Figs. \ref{transverse_res}(c)-(f). Specifically, this becomes apparent at z = 0 nm for the 359 nm resonance where the rods interact strongly [Fig. \ref{transverse_res}(c)] and at higher z-coordinates for the 369 nm resonance which is represented at z = 10 nm [Fig. \ref{transverse_res}(f)].

Meanwhile, $\mathscr{S}_{scat, forward}$ at 347 nm is negative and of significantly smaller magnitude [bottom panel of Fig. 3(b)]. Again, this corresponds to our observations of the near-field optical chirality density shown in Figs. \ref{transverse_res}(a) and (b).

\subsection{Circular Dichroism}
\begin{figure}
\includegraphics{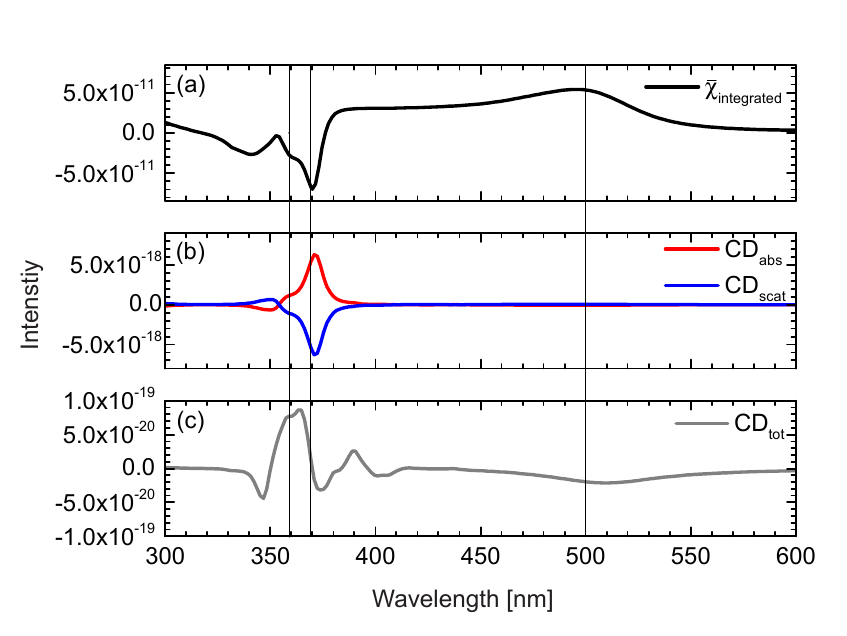}
\caption{Spectral information on the Ag chiral plasmonic nanorod dimer studied in Sec. \ref{sec_NR}. (a) Integrated optical chirality density $\bar{\chi}$ in newtons in the space surrounding the structure. (b) Absorbed ($CD_{abs}$) and scattered ($CD_{scat}$) components of the CD spectrum in watts. (c) Total CD spectrum $CD_{tot} = CD_{abs} + CD_{scat}$ in watts.}
\label{CD_1}
\end{figure}

Figure \ref{CD_1} provides additional information on the cancellation effects which occur in the circular dichroism (CD) spectra of the chiral nanorod dimer.  At resonance, the interaction between the nanorods leads to a redistribution of charges within each rod. In the dipole limit, the resulting shift in the field lines causes a reorientation of the dipole moment of each nanorod. Thus, the coupled dipole oscillator model \cite{fan2010plasmonic, norden1997circular} can explain the observations made in the CD spectra shown here.

Figure \ref{CD_1}(b) shows that for this structure, the absorption and scattering components of the CD spectrum are of similar magnitude and opposite in sign. This effect can also be described by the coupled dipole oscillator model \cite{fan2010plasmonic, norden1997circular} and represents an additional source of undesirable cancellation effects, which can arise from the detection of the sum of absorbed and scattered components in an extinction measurement. 
The means by which these cancellation effects can skew the resulting spectral information is demonstrated by comparing Fig. \ref{CD_1}(a), which shows the integrated optical chirality density surrounding the structure [identical to Fig. \ref{figure_3}(b)], and Fig. \ref{CD_1}(c), which shows the sum of the absorbed and scattered components of the CD spectrum. The latter corresponds to a CD extinction measurement. Although when considered separately, the resonances of the absorbed and scattered components of the CD spectrum match those of the integrated optical chirality density, they are skewed and shifted due to cancellation effects in the total CD spectrum [Fig. \ref{CD_1}(c)].

\end{document}